# Identification of cancer-keeping genes as therapeutic targets by finding network control hubs

Xizhe Zhang[1,2*], Chunyu Pan[1], Xinru Wei[1,2], Meng Yu[3,4], Shuangjie Liu[5], Jun An[5], Jieping Yang[5], Baojun Wei[5], Wenjun Hao[5], Yang Yao[6], Yuyan Zhu[5*], Weixiong Zhang[7,8*]

1. Early Intervention Unit, Department of Psychiatry, Affiliated Nanjing Brain Hospital, Nanjing Medical University, Nanjing, China
2. School of Biomedical Engineering and Informatics, Nanjing Medical University, Nanjing, China
3. Department of Laboratory Animal Science, China Medical University, Shenyang, China.
4. Key Laboratory of Transgenetic Animal Research, China Medical University, Shenyang, China.
5. Department of Urology, First Affiliated Hospital of China Medical University, Shenyang, China
6. Department of Physiology, Shenyang Medical College, Shenyang, China
7. Department of Health Technology and Informatics, The Hong Kong Polytechnic University, Hong Kong
8. Department of Computer Science and Engineering, Department of Genetics, Washington University in St. Louis, St. Louis, MO, USA

*: Correspondence. XZ: zhangxizhe@gmail.com; YZ: yyzhu@cmu.edu.cn;
WZ: weixiong.zhang@polyu.edu.hk

**Abstract**

Finding cancer driver genes has been a focal theme of cancer research and clinical studies. One of the recent approaches is based on network structural controllability that focuses on finding a control scheme and driver genes that can steer the cell from an arbitrary state to a designated state. While theoretically sound, this approach is impractical for many reasons, e.g., the control scheme is often not unique and half of the nodes may be driver genes for the cell. We developed a novel approach that transcends structural controllability. Instead of considering driver genes for one control scheme, we considered control hub genes that reside in the middle of a control path of every control scheme. Control hubs are the most vulnerable spots for controlling the cell and exogenous stimuli on them may render the cell uncontrollable. We adopted control hubs as cancer-keep genes (CKGs) and applied them to a gene regulatory network of bladder cancer (BLCA). All the genes on the cell cycle and p53 singling pathways in BLCA are CKGs, confirming the importance of these genes and the two pathways in cancer. A smaller set of 35 sensitive CKGs (sCKGs) for BLCA was identified by removing network links. Six sCKGs (RPS6KA3, FGFR3, N-cadherin (CDH2), EP300, caspase-1, and FN1) were subjected to small-interferencing-RNA knockdown in four cell lines to validate their effects on the proliferation or migration of cancer cells. Knocking down RPS6KA3 in a mouse model of BLCA significantly inhibited the growth of tumor xenografts in the mouse model. Combined, our results demonstrated the value of CKGs as therapeutic targets for cancer therapy and the potential of CKGs as an effective means for studying and characterizing cancer etiology.

**INTRODUCTION**

Whole-genome and exome sequencing has enabled the identification of cancer driver mutations and cancer driver genes (CDGs)[1, 2], which are introduced to elucidate cancer etiology[3] and help identify putative therapeutic targets for cancer treatment[4]. Most existing methods for CDG discovery are in essence frequency-based[5], some of which are enhanced by additional functional analysis[6]. They look for somatic mutations that are statistically significant for separating cancer subjects and normal controls and are thus regarded as putative drivers of cancer development[7, 8]. Much effort, e.g., The Cancer Genome Atlas (TCGA) project[1], has been devoted to identifying CDGs[2, 9] and large quantities of data have been collected. However, CDGs that can be discovered seem to be saturated even with increasing sample sizes[10] and few CDGs have been validated by independent studies. Critically, the existing CDG-finding methods are in principle association-based that correlate individual genetic mutations in isolation with disease phenotypes; as such, they produce candidate driver and passenger mutations altogether[11]. Identifying genuine CDGs from a large pool of candidates remains a challenge.

Apart from the frequency-based methods is an approach based on network structural controllability[12, 13]. Network controllability characterizes how a networked system can be driven from a state to the desired state by external stimuli on the driver nodes of a control scheme for the network[12]. Network controllability has been applied to various biological networks[14-16], including protein-protein interaction networks[17], gene regulatory networks[18, 19], and metabolic networks[20, 21]. Instead of analyzing individual genes in isolation, this approach organizes all genes of interest in a network and analyzes them as a whole to identify driver nodes or genes. When applied to cancer regulatory networks, it identifies driver nodes that are regarded as CDGs and therapeutic targets for precision cancer treatment[22]. However, two critical issues must be addressed to make network controllability effective for finding CDGs. Firstly, given a network, the control scheme is often not unique, but rather numerous control schemes and different sets of driver nodes exist[23, 24]. It is undetermined which control scheme can most effectively control the network. One may compare all control schemes to select the best, e.g., one with the shortest control path to the desired state. However, finding all control schemes, a #P-hard problem[25], is computationally prohibitive. In addition, one control scheme may contain a substantial number of driver genes[23] which need to be subjected to exogenous stimuli together to change the state of the cell. It is impractical to manipulate a large number of genes for disease treatment. Secondly and more importantly, a fundamental, albeit subtle and mostly neglected, assumption underlying network controllability based methods is that the biological network being analyzed is a model describing both cancerous and normal states of the cell so that mutating some putative CDGs can drive the cell from a normal state to a cancerous state, resulting in cancer. However, little is known about which states are normal and which other states are cancerous, so a driver node in the network may not necessarily be a CDG. Therefore, it is challenging to make the network controllability theory a practical means for identifying CDGs and cancer therapeutic targets.

We aim at extending network controllability to an effective approach for cancer research and treatment. We focus on homogenous networks that model either the normal cell or a cancer cell, and importantly go beyond one control scheme and consider the overall network controllability governed by all control schemes of a network. We introduce *control hubs*, which are hub nodes that belong to all control schemes, and *sensitive control hubs* that are sensitive to small structural perturbations. We developed a polynomial-time algorithm for finding all sensitive control hubs without computing all control schemes. When applied to a cancer regulatory network, sensitive control hubs lend themselves to *sensitive cancer-keeping genes* (sCKGs) that maintain the validity of the cancer regulatory network. We hypothesized and argued that sCKGs were excellent candidates for cancer biomarkers and therapeutic targets. We applied our approach to the bladder cancer (BLCA) gene regulatory network and identified 35 sCKGs for the disease. Using small-interferencing RNA (siRNA) knockdown, we validated six sCKGs *in vitro* and confirmed their effects on the proliferation of four cancer cell lines. We also validated an sCKG *in vivo* in a mouse model of BLCA by showing its prohibitive function on tumor progression.

**RESULTS**

*Control hub nodes of a network*

A key observation that pillars structural controllability is that a node in a directed network can control one of its outgoing neighbors[26] and the controlled neighbor can control a neighbor of its own and so on. These nodes collectively form a *control path* that has a head node at the beginning of the path, also referred to as *control node*[12] or *driver node*[13], a tail node at the end of the path, and middle nodes if any (Figures 1A and 1B, Supplemental eMethod 1 and 2). All driver nodes and their control paths constitute a *control scheme* of the network (Figure 1B). It is important to note that the control scheme is usually not unique for a complex network[23, 24]. A node may take different positions and thus play distinct roles in different control schemes (e.g., nodes *c* and *d* in Figure 1B). It is difficult to determine the best control scheme. We may compute all control schemes to select the best. However, it is computationally prohibitive to derive all control schemes because the problem is #P-hard[25], meaning that no polynomial algorithm is known for the problem. Furthermore, given all control schemes, it is still nontrivial to determine the best one because little is known about the network dynamics and different optimality criteria (e.g., the minimal set of control nodes versus the fewest steps to reach the desirable state) will lead to distinct control schemes. It is also important to note that a substantial number of nodes may serve as driver nodes in different control schemes[23], making it costly to control the network. All these unfavorable factors cast doubts on the utility of a direct application of network structural controllability.

We go beyond one control scheme and consider the overall structural controllability of a network by considering all control schemes. To avoid the excessive computational burden, we do not compute all control schemes but instead, focus on all *control hubs* that are middle nodes of some control paths of *every* control scheme (e.g., node *e* in Figure 1B). An eminent feature of a control hub is that it is essential for maintaining the overall control structure of the network – changing it to a head or tail node in any control scheme will void all control schemes and consequently take apart the overall control structure of the network. Therefore, it is critically important to protect all control hubs to maintain network controllability and the validity of the network model. This implies that control hubs are the most vulnerable spots for network controllability. Here, we explicitly explore and exploit this property of control hubs.

However, identifying all control hubs of a large network is technically nontrivial. We developed a novel polynomial-time algorithm for finding all control hubs without computing all control schemes[27]. The algorithm is based on previous work on enumerating all driver nodes[23, 28]. It first identifies the head and tail nodes of the control paths of all control schemes and subsequently identifies control hubs. It has a complexity of $O(n \times m)$ on a network with $n$ nodes and $m$ edges.

*Control hub nodes as cancer-keeping genes*

In a mixed model representing both the normal and cancerous states of the cell, it is difficult or even infeasible to discern if a state represents the normal state or a cancerous state. Likewise, it is difficult to determine if exogenous stimuli can transfer the cell from the normal state to a cancerous state or vice versa.

To address this issue, we consider the homogenous model that represents exclusively the normal cell or a cancerous cell and focus on the overall network controllability using control hubs. When applied to a cancer regulatory network (a model of a cancerous cell), control hubs are preferred over driver nodes to be actionable targets for cancer treatment. When exogenous stimuli turn a control hub into a non-control hub, the network is no longer controllable by any control scheme. As a result, the cell must have transitioned out of the current cancerous state to, presumably, the normal state. Therefore, we name control hub genes in a cancer regulatory network *cancer-keeping genes* (CKGs) because they maintain the network controllability of the model. Furthermore, we suggest adopting CKGs as actionable targets for cancer treatment. We experimented with this idea using cancer cells and a mouse model of bladder cancer, to be discussed shortly.

When used to control a network, the fewer control hubs, the better. It is particularly true with cancer gene regulatory networks on which potential drug targets should be kept as few as possible. While a network typically has fewer control hubs than driver nodes, the control hubs may still be more than what we can experimentally manipulate. We suspect if two control hubs are created equal. We expect that some control hubs are more sensitive and vulnerable to external perturbations than other control hubs. We are particularly interested in those control hubs that can be turned into non-control hubs when a single edge is removed from the network as a perturbation (Figure 1C), which we call *sensitive control hubs* or *sensitive CKGs* (sCKGs). All sCKGs can be identified by removing every edge of the network one at a time (see Methods).

***Cancer-keeping genes are essential in the gene regulatory network of bladder cancer***

We applied our novel CKG approach to bladder cancer (BLCA). We first introduced a novel method to construct a regulatory network for BLCA, short-handed as BLCA_GRN (Figure 2A, see Methods). BLCA_GRN was built using the data from 45 known BLCA driver gene[29], 50 most mutated genes in BLCA based on TCGA data (Supplemental eTable 1), and the data from five manually curated interaction databases (Supplemental eTable 2). The resulting BLCA_GRN has 7,030 nodes or genes and 103,360 directed edges or interactions (Figure 2B, Supplemental File 1). One control scheme has 3,115 driver genes (44.3% of all 7,030 genes) and there exist 5,052 driver genes (71.9%) for all control schemes (Figure 2C). So many control schemes and nearly two-thirds of all genes being driver genes made it difficult, if not infeasible, to choose the right control scheme for BLCA when applying structural controllability.

Our new CKG approach identified 660 CKGs in BLCA_GRN, which are 9.4% of all 7,030 genes and 13.1% of all 5,052 driver genes in the network (Figure 2C). Fewer CKGs made them a better choice as actionable targets for treating BLCA. The 660 CKGs have several characteristics. First, they have greater connectivity in the network, with an average degree of 96.4, than the rest of the nodes that have an average degree of 29.4 (Figure 2D), reflecting the greater controlling power of the CKGs. Next, we characterized CKGs, head and tail genes in the context of essentiality, evolutionary conservation, and regulation at the levels of translational and posttranslational modifications (PTMs) as done previously[30]. The results showed that the CKGs were significantly enriched in most of the above datasets. (Figure 2E, Supplemental eFigures 2-4). More interestingly, the CKGs were significantly enriched with the targets of pathogenic mutants, human viruses, drug targets, and immunotherapy targets (Figure 2E, Supplemental eFigures 5-6). Altogether, these results revealed that CKGs play crucial regulatory roles.

Furthermore, comparing the potential targets of tumor immunotherapy including core cancer-intrinsic CTL-evasion genes (coreCTL)[31], CAR therapy targets[32], and immune checkpoints[33, 34] showed that 27 CKGs were known targets of tumor immunotherapy, in which 20 CKGs were coreCTL genes and 7 CKGs were CAR genes or resided at classical immune checkpoints. Surprisingly, although 437 CKGs were not directly immune-related, their neighboring genes in the network were tumor immune regulatory genes (Supplemental File 2). Combined, these results strongly suggested that the 660 CKGs were all involved in multiple regulatory processes of tumor immunity, reflecting the functional importance of 660 CKGs in the networked control of cancer.

To assess the potential oncogenic functions of the 660 CKGs, we examined their involvement in ten key cancer-related signaling pathways that have genes with extensive genetic variations in 33 types of cancer as analyzed by TCGA[35]. Interestingly, 70.4% of the genes in four cancer signaling pathways (i.e., the cell cycle, p53, TGFß, and RTK-RAS pathways) were CKGs (Figure 2F, Supplemental eTable 4). Strikingly, all the genes in the cell cycle and p53 pathways were CKGs. Many CKGs (e.g., the CKG ACVR1B in the TGFß pathway and the CKGs (FGFR3, KIT, NTRK2, and BRAF) in the RTK-RAS pathway) were located upstream of the pathways. These results indicated that CKGs were tightly regulated and played critical roles in cancer, particularly BLCA.

*Sensitive cancer-keeping genes have oncogenic functions and clinical importance in BLCA*

To select a small subset of the 660 CKGs as potentially druggable targets[36] for treating BLCA, we looked for sCKGs in BLCA_GRN. The rationale for taking edge removal as network perturbation is that many cancer drugs block protein-protein interactions, thus acting as edge removal. One example is an irreversible Fibroblast growth factor inhibitor Futibatinib[37] for BLCA. We name an edge *sensitive edge* if removing it renders a CKG to become an sCKG. An sCKG has at least one sensitive edge associated; the more sensitive edges that an sCKG is associated with, the more sensitive and druggable it is. Of the 660 CKGs, 173 (26.2%) were sCKGs, among which more than 35.8% were associated with more than one sensitive edge (Figure 3A). All sensitive edges had high confidence scores[38] (Supplemental eTable 5), indicating that it is highly confident that the sCKGs were sensitive to such external stimuli. The 173 sCKGs have several characteristics. They predominantly had moderate or low connectivity; their average connectivity was less than half of that of the other CKGs (Figure 2D). Counter-intuitively, the sCKGs also had significantly lower rates of genetic alteration (including mutation, copy number variation, or homozygous deletion) than their neighbors in BLCA_GRN (Figure 3B, Supplemental File 2), so they may not be detected by frequency-based methods.

We mapped the 173 sCKGs to the indispensable genes of PPI network[30] to produce a small set of 35 sCKGs that might be potentially actionable therapeutic targets (Figure 3D). Most of the sCKGs were directly connected to the known cancer driver genes[29] and/or cancer therapeutic targets[39] in BLCA_GRN (Figures 3C and 3D, Supplemental eFigure 7) and were enriched in functional and regulatory genes datasets (Supplemental eFigure 8). For example, 26 of the 35 sCKG had cancer driver genes in their direct neighbors (Figure 3D, Supplemental eTable 6). In addition, 30 (85.7%) of the 35 sCKGs had cancer therapeutic targets in their direct neighbors (Figure 3D, Supplemental eFigure 9B, and eTable 7).

Interestingly, many of the 35 sCKGs had epistatic relations. Specifically, increased mutations to more than one of these genes significantly increased the overall survival rates of many BLCA subjects based on the TCGA PanCancer clinical data (see Methods). Comparing the median survival times of all BLCA patients with two or more of the 35 sCKGs being mutated revealed that the median survival time would increase from 28.41 months (for the unaltered group) to 59.31 months (for the one sCKG altered group), 66.41 months (for the two sCKGs altered group), and 104.65 months (for the three and more sCKGs altered group) under the stringent criterion of the Logrank Test P-values being less than 0.015 (Figure 3E). This result strongly suggested that these genes were putative diagnostic and prognostic biomarkers and potential therapeutic targets for BLCA.

Two sCKGs (FGFR3 and RPS6KA3) deserved further scrutiny. They are well-characterized CDGs for BLCA[6] and are known to involve in two key signaling pathways of BLCA, RTK-RAS-PI(3)K and p53[40] (Figure 2G). Remarkably, FGFR3 had PIK3CA as a downstream gene in the RTK-RAS-PI(3)K pathway and RPS6KA3 was upstream of TP53 in the p53 pathway, suggesting that FGFR3 and RPS6KA3 were upstream drivers of two key CDGs. Furthermore, RPS6KA3 is a substrate of FGFR3 so they are related and share many common functions. FGFR3 can phosphorylate Y529 and Y707 of RPS6KA3 to assist ERK1/2 to connect to RPS6KA3 and keep RPS6KA3 active[41]. Note that the interaction between FGFR3 and RPS6KA3 was critical. The removal of the link between the two genes in BLCA_GRN would change the two genes from control hubs to head nodes and subsequently invalidate all control schemes for BLCA_GRN, suggesting that the two genes must play critical roles in BLCA.

Immunotherapy has been adopted in treating BLCA and several drugs (e.g., Atezolizumab and Avelumab) have been developed to target immune checkpoint inhibitors (e.g., PD-1 and PD-L1)[42]. Among the six well-known genes targeted by BLCA drugs, five (CD274/PD-L1, CTLA4, IL12B, PTGS2, and TFDP1) were head or tail nodes except FKBP1A (Supplemental eTable 9). Nevertheless, all six genes were well connected with CKGs (i.e., control hubs) in BLCA_GRN. For example, CTLA4 has seven neighbors and six of them were CKGs. Such a close connection with CKGs supported taking these genes as potential drug targets for BLCA treatment.

*Cancer-keeping genes as potential therapeutic targets for bladder cancer*

We subjected a handful of sCKGs for BLCA to *in vitro* and *in vivo* analyses to assess their impact on cancer cell proliferation and migration, thus assessing and confirming their function in maintaining cell states and viability. We adopted the technique of small interference RNAs (siRNAs) to knock down six sCKGs (RPS6KA3, FGFR3, N-cadherin (CDH2), EP300, caspase-1, and FN1) in four cancer cell lines and an sCKG (RPS6KA3) in a mouse model of BLCA. When choosing sCKGs for experimental analysis, we avoided known CDGs and genes that have been well characterized for BLCA. The cell lines considered represent three types of cancer, i.e., T24 and UMUC3 for BLCA, SiHa for cervical cancer, and FaDu for head-and-neck cancer. The siRNA knockdown experiments were repeated three times for every gene and every cell line and the experiments were also repeated on ten mice, for a total of 42 siRNA cell-line assays and 10 mouse experiments. For each experiment, the expression of the sCKG was measured by quantitative PCR. The results of the repeated experiments were consistent (Figure 4A).

The absence of RPS6KA3, FGFR3, and CDH2 significantly decreased the proliferation of all four tumor cells (Figures 4A and 4B; Supplemental eFigures 10A and 10B), whereas the knockdown of EP300 and FN1 significantly promoted the proliferation of bladder UMUC3 tumor cells (Figure 4B). Interestingly, the loss of caspase-1, which was thought to play an important role in inflammation, promoted the proliferation and survival of BLCA cells (Figures 4A and 4B). Moreover, the deletion of FGFR3, RPS6KA3, EP300, FN1, and CDH2, as expected, significantly reduced the migration ability of bladder tumor cells (Figures 4C and 4D). The role of CDH2 in promoting cell migration was also confirmed in head-and-neck cancer cells (Supplemental eFigure 10B). In contrast, the absence of caspase-1 promoted the migration of BLCA cells (Figures 4C and 4D).

Furthermore, we used a nude mouse model with transplanted BLCA tumors to examine the function of RPS6KA3, which can promote the proliferation of many tumor cells. Consistent with the *in vitro* results, the knockdown of RPS6KA3 in the mouse model significantly inhibited the growth of BLCA xenografts in the mice (Figure 4E). Putting together, all the *in vitro* and *in vivo* experimental results confirmed the functional roles of some of the sCKGs for maintaining the validity of cancer cells and validated the feasibility of our novel network control hub-based method.

**DISCUSSION**

The well-developed theory of network controllability[13] provides a novel perspective on the dynamics of biological networks and how critical genes, particularly CDGs and cancer therapeutic targets, can be identified[30, 43]. It is fundamentally different from the conventional mutation-based approaches for finding CDGs. Although theoretically sound, however, network controllability does not offer an effective method for finding CDGs or drug targets, as discussed earlier.

In the current research, we developed a novel approach that transcended network controllability and importantly transformed the latter into a feasible and effective means for identifying potential biomarkers for cancer diagnosis and putative drug targets for cancer treatment. Our approach hinged upon three ideas atop network controllability. First, network controllability should be applied to homogenous networks that represent exclusively cancerous cells or the normal cell to avoid the issue of determining if the current network state represents a normal or cancerous cell. Second, making a network *uncontrollable* by any of its control schemes may capture the migration of the normal cell to a cancerous cell or vice versa. Third, the control hubs that we introduced are the most vulnerable components of the network that must be protected to maintain network controllability. Therefore, targeting control hubs may potentially change the current cell state and properties. In this research, we capitalized on the notion and properties of control hubs and introduced cancer-keeping genes (CKGs) by applying control hubs to cancer gene regulatory networks.

Importantly, there are typically much fewer control hubs (or CKGs) than driver nodes in a gene regulatory network. For example, BLCA_GRN has 5,052 driver genes out of a total of 7,030 genes but only 660 CKGs can be further reduced to 173 sCKGs. Most CKGs for BLCA are involved in important tumor signaling pathways. Remarkably, the genes in the cell-cycle and p53 signaling pathways as reported by the TCGA project[35] are all CKGs (Figure 2F). Dysregulation of cell-cycle control is a hallmark of cancer[44]. Many CKGs in the cell-cycle pathway, including CDKN1A[45], CDK2[46], and E2F1[47], have been considered potential targets of anticancer drugs. The p53 pathway is tightly regulated in BLCA[48] and the genetic variations of the genes in the pathway have been an attractive topic for BLCA treatment[49, 50]. CDKN1A and TP53 in the cell-cycle and p53 pathways, respectively, have been taken as therapeutic targets for BLCA[51]. The enrichment of the CKGs in TGFß and receptor-tyrosine kinase (RTK)/RAS/MAP-Kinase signaling pathways is also thought-provoking. The two pathways are known to be important for BLCA. TGFß can act as a critical tumor suppressor[52] and the dysregulation of the TGFß pathway may increase the risk of BLCA. Sixteen CKGs are also involved in other cancer-related signaling pathways, including PI3K, Myc, Wnt, Notch, Hippo, and Nrf2 signaling pathways (Supplemental eTable 4).

We like to highlight that CKGs are network-structure-based so that they are fundamentally different from mutation-based CDGs. Most known CDGs have high mutation rates and/or high connectivities in gene regulatory networks. In contrast, CKGs, particularly sCKGs, have average mutation rates and thus are unlikely to be detected by the existing mutation-based methods. On the other hand, most CKGs are directly connected to known CDGs and drug targets in BLCA_GRN, forming gene pairs that are essential for the maintenance and function of the network. Many CKGs are also located upstream of tumor-related signaling pathways and thus can control CDGs. Therefore, CKGs provide alternatives to CDGs, and many CKGs can be viewed or adopted as latent, master regulators of CDGs. Indeed, manipulating CKGs can affect cancer cell proliferation and migration and suppress tumor growth, as we have shown in our experiments using four cancer cell lines and a mouse model of BLCA. The six sCKGs (RPS6KA3, FGFR3, N-cadherin (CDH2), EP300, caspase-1, and FN1) that we experimentally analyzed are neither known CDGs nor drug targets and have not been well studied. Combined, our analytic and experimental results strongly suggest CKGs be a class of novel regulatory elements that, when perturbed, can affect the states of the underlying cells. In particular, the six experimentally analyzed sCKGs are novel putative drug targets for BLCA treatment.

Although we introduced CKGs for cancer-cell models, the underlying idea of control hub is general and the idea and reasoning can be equally applied to the normal cell for identifying control hubs as *cancer-causing genes*. The extension to network controllability as done in our work is thought-provoking and is expected to inspire future work to extend the theory of network controllability in multiple dimensions for the urgent demands for innovative methods for analyzing large quantities of biological data. Our novel control-hub-based analytic approach can be directly applied to various types of cancer and be readily extended to other complex diseases, such as neurodegenerative disorders and metabolic diseases.

**MATERIALS and METHODS**

*Construction of bladder cancer gene regulatory network*

We developed a novel method for constructing cancer gene regulatory networks consisting of control-related interactions and cancer-related genes. We considered the ten most important and common types of control-related interactions (Supplemental eTable 2) from five well-curated pathway databases, including the National Cancer Institute (NCI) Nature Pathway Interaction Database[53], PhosphoSite Kinase-substrate information[54], HumanCyc[55] (https://humancyc.org), the Reactome[56] (https://www.reactome.org) and PANTHER Pathways[57] (http://www.pantherdb.org). From the data generated by the TCGA Research Network (https://www.cancer.gov/tcga), we chose 45 known BLCA cancer driver genes[29] and 50 most mutated genes of BLCA as seed nodes (Supplemental eTable 1). A breadth-first search, starting from the seed nodes, was adopted to traverse the control-related interactions. The traversed genes and the interactions formed the BLCA gene regulatory network (BLCA_GRN) with 7,030 nodes (genes) and 103,360 directed edges (interactions) (Supplemental File 1).

*Identification of control scheme*

To find a control scheme for BLCA_GRN, network $G(V, E)$ was first converted to an equivalent undirected bipartite graph $B(V_{in}, V_{out}, E)$ by splitting the node-set $V$ into two sets $V_{in}$ and $V_{out}$. Here, a node $n$ in $G$ is converted to two nodes $n_{in}$ and $n_{out}$ in $B$, where $n_{in}$ and $n_{out}$ are, respectively, connected to the in-edges and out-edges of node $n$. Following an early work[58], a minimum set of driver nodes $D$ can be obtained by computing a maximum matching using the Hopcroft-Karp algorithm[59]. The unmatched nodes are the driver nodes. The driver nodes and the corresponding edges in the maximum matching form a control scheme.

The driver nodes of all control schemes can be obtained based on our previous work[28].

*Identification of control hub nodes or cancer-keeping genes*

For a directed network $G(V, E)$, we say a node is a *head node* or *tail node* if it is such a node in any control path of any control scheme of the network. A node is a *control hub node* (or *control hub*) if it resides in the middle of a control path in *every* control scheme of $G(V, E)$. Based on the above definition, a node is a control hub *iff* it is not a head or tail node. Therefore, the *control hub nodes* can be identified by our previous works[27], which are listed in the following steps:

Step1. Find the set of all driver nodes $H$ by the algorithm of our previous work[28];
Step2. Construct the transposed network $G^T(V, E^T)$ of $G(V, E)$ by reversing the direction of every edge;
Step3. Find the set of all driver nodes $T$ of $G^T(V, E^T)$; The set $V\setminus(H\cup T)$ is the set of control hubs of $G(V, E)$.

The main component of the algorithm is finding all driver nodes with a complexity of $O(n^{0.5}m)$ on a network with $n$ nodes and $m$ edges[24], and so is the complexity for finding all control hubs.

*Identification of sensitive control hub nodes or sensitive cancer-keeping genes*

A control hub is referred to as being *sensitive* if it is no longer a control hub after removing any edge from the network. Therefore, the sensitive control hubs of a network can be identified by removing edges one at a time and examining whether the set of control hubs of a control scheme remained intact. The complexity of this algorithm is O($n^{0.5}m^2$) on a network with $n$ nodes and $m$ links in the worst.

*Functional enrichment analysis of cancer-keeping genes*

We adopted the enrichment analysis method of previous work[30] by computing the significance of the overlap between a set of nodes and a given functional dataset. The node set included CKGs, head genes, and tail genes. A total of 32 function databases (Supplemental eTable 3) were adopted to assess the functional significance of the above three node sets.

*Map sensitive CKGs with cancer driver genes and cancer therapeutic targets*

The CKGs were further analyzed using several datasets including that of CDGs, cancer therapeutic targets, and immune genes. The CDG dataset[29] includes 739 genes obtained based on 9,423 tumor exomes from TCGA PanCancer analysis with 26 computational tools to catalog driver genes and mutations. Cancer therapeutic target dataset[39] includes 628 genes obtained using genome-scale CRISPR-Cas9 screening of 324 human cancer cell lines from 30 cancer types. Immune-related genes include 358 genes which were obtained by coreCTL immune-related genes[30], CAR immune-related gene datasets[31], and immune checkpoints[33, 34]. We took the overlapped genes between the three datasets with the CKGs for further analysis.

*Survival analysis*

We performed a survival analysis of 35 sCKGs using the clinical data of the 411 samples of Bladder Urothelial Carcinoma (TCGA PanCancer analysis) from the cbioportal website (http://www.cbioportal.org). We counted the number of mutated genes in 35 sCKG across all 411 samples and selected 4 subsets based on the number of simultaneously mutated genes. Whether a gene was considered mutated depends on the common alterations (somatic mutation, gene fusion, copy number amplification, or homozygous deletion) it contained. These subsets respectively had no mutated genes, at least one mutated gene, at least two mutated genes, and at least three mutated genes in the 35 sCKGs. The survival analysis was performed by using the Kaplan-Meier estimator[60] from the cbioportal website (http://www.cbioportal.org).

*Cell culture*

Cancer cell lines were cultured in RPMI-1640 media (T24, UMUC3) and MEM media (FaDu and SiHa) supplemented with 10% FBS. All cell lines were cultured in the atmosphere of 5% $CO_2$ at 37°C and verified to be mycoplasma negative. Lipofectamine 3000 (Invitrogen) was used to transfect cancer cell lines.

*Transient siRNA-mediated gene knockdown*

The negative control (siRNA-NC) and siRNA targeting candidate CKGs were purchased from Shanghai Genechem Co., Ltd. (China). siRNA was diluted to a final concentration of 20 μM following the manufacturer's instructions. Cancer cells were transfected using Lipofectamine TM RNAiMAX (Invitrogen, Carlsbad, CA, US) and opti-MEM (Gibco). Twenty-four hours after transfection, cells were collected and subjected to the subsequent experiments. The sequences of the siRNA were listed in Supplemental eTable 10-11.

*Cell Counting Kit 8 (CCK8) assay*

The effect of knocking down candidate CKGs on cell proliferation was examined by CCK-8 assays. Twenty-four hours after transfection, cells were collected and maintained in 96-well plates (1,500 cells/well). Four hours after incubation, 10 μl of CCK-8 reagent (Vazyme, Nanjing, China) was added to each well and the reaction mixtures were incubated for 2 h at 37 °C under a 5% $CO_2$ atmosphere. The absorbance (OD) value at 450 nm wavelength was then measured using a Microplate reader (Bio-Rad, CA).

*Transwell assay*

Transwell assays were employed to determine the effect of candidate cancer-keeping genes on the migration and the invasion abilities of cancer cells. Twenty-four hours after transfection, cells were collected and subjected to transwell assay. In brief, the upper compartment of the filters (Corning, NY, USA) was coated with (invasion) or without (migration) 55 µl of the basal membrane matrix (1:7 dilution, Corning). Cells were diluted with serum-free medium to 1× $10^5$ cells per ml. Cell suspension (200 µl) was added to the upper compartment, and 600 µl of RPMI-1640 containing 10% FBS was added to the lower compartment. Twenty-four hours after incubation at 37°C under 5% $CO_2$, filters were fixed with 4% formaldehyde for 30 min, then stained with 0.1% crystal violet at room temperature for 30 min. Filters were then rinsed 3 times with PBS, and the unmigrated cells were removed with cotton swabs. Finally, cells were photographed and counted in a 5-independent microscopic field.

*Xenograft mouse model*

UMUC3 cells expressing control shRNA or PRS6KA3 shRNA (2×$10^6$) were subcutaneously injected into the dorsal flank of 4-week-old male athymic nude mice (*n*=5 mice per group, Shanghai SLAC Laboratory AnimalCo. Ltd.). Mice were sacrificed after 3 weeks and tumors were excised and weighed. Mice were used in the experiment at random. During testing the tumors' weight, the researchers were blinded to the information and shape of tumor tissue masses. Studies on animals were conducted with approval from the Animal Research Ethics Committee of the China Medical University.

**Data and software availability**

The software of our method for finding control hub nodes and cancer-keeping genes is freely available in the public software repository github at https://github.com/network-control-lab/control-hubs.

**Disclosure of Potential Conflicts of Interest**

No potential conflicts of interest were disclosed.

**Author contributions**

Xizhe Zhang conceived and designed the project as well as supervised the work except the biological experiments. Xizhe Zhang also designed the algorithm, analyzed its correctness and complexity, performed data analysis, and draft the manuscript. Weixiong Zhang conceived the main conceptual ideas and draft the manuscript. Chunyu Pan and Xinru Wei implemented the algorithm and collected and analyzed the data. Yuyan Zhu designed and supervised the *in vivo* and *in vitro* experiments, and performed data analysis and integration and related manuscript writing. Yu Meng was involved in the design of *in vivo* experiments and related data analysis. Shang Liu, Jun An, and Jieping Yang carried out the *in vitro* and *in vivo* experiments and data collection. Baojun Wei and Wenjun Hao analyzed the data and results from the *in vitro* and *in vivo* experiments.

**Acknowledgment**

This work was supported in part by the National Natural Science Foundation of China (grant numbers 62176129 and 81672523), the United States National Institutes of Health (grant number R01-GM100364), and the Hong Kong Global STEM Professorship Scheme.

**ORCID**

Xizhe Zhang: http://orcid.org/0000-0002-8684-4591; Yuyan Zhu: http://orcid.org/0000-0003-0479-3151; Weixiong Zhang: http://orcid.org/0000-0002-4998-9791

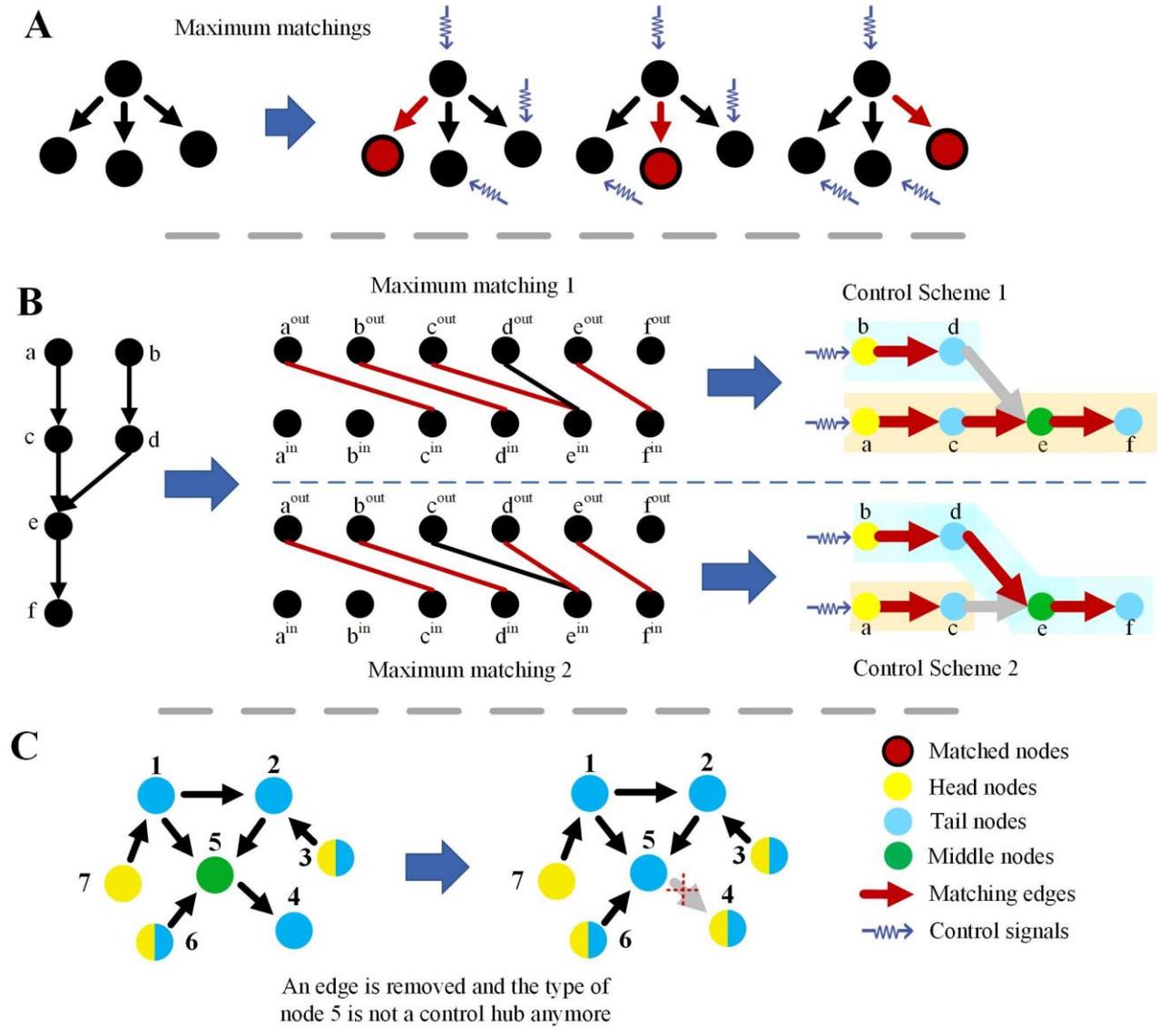

**Figure 1** | Control hubs of complex networks. **A)** An example of a simple network with three different maximum matchings. The unmatched nodes are driver nodes and the red edges are matched edges. **B)** Two different control schemes of a simple network. A node may lie at the head, tail, or middle of a control path. Node *e* resides in the middle of some control paths of the two control schemes, becoming a control hub of the network. **C)** Sensitive control hubs of a simple network. After removing edge *e*(5,4), node 5 changes from a control hub to a tail node, therefore, it is a sensitive control hub of the network.

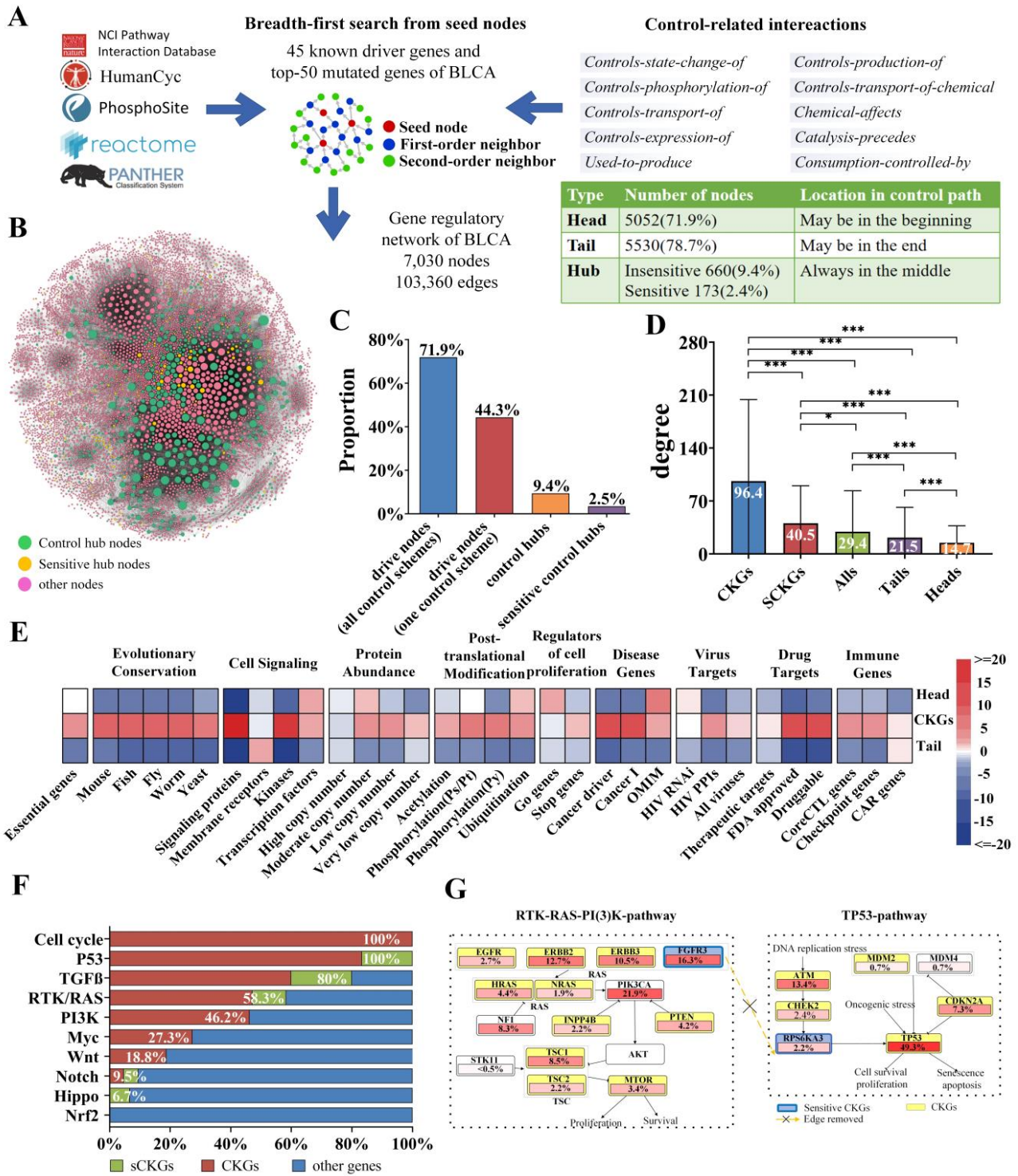

**Figure 2 |** Control hubs or cancer-keeping genes in the gene regulatory network of bladder cancer (BLCA_GRN). **A)** Construction of the BLCA_GRN. We first extracted ten types of control-related gene/protein interactions from five high-quality pathway databases. We then performed a breadth-first search starting from 45 known BLCA driver genes and the 50 most highly mutated genes of BLCA. The traversed nodes and edges constitute the BLCA_GRN. **B)** Topological structure of the BLCA_GRN. Node size is proportional to node degree. The network contains 7,030 genes and 103,360 directed interactions. **C)** Proportions of the driver nodes, control hubs, and sensitive control hubs in the BLCA_GRN. **D)** Node degrees of different types of genes in BLCA_GRN. Control hubs have a significantly higher average degree than the other genes. We used one-way ANOVA (analysis of variance) followed by multiple comparison tests (post-hoc) with a 0.05 significance level. ***p < 0.001; **p < 0.01; *p < 0.05. **E)** Control hubs of

BLCA_GRN are enriched in the context of essentiality, evolutionary conservation, cell signaling, protein abundance, post-translational modifications (PTMs), regulators of cell proliferation, diseases, virus targets, drug targets, and immune regulation. **F**) Proportions of sensitive CKGs and CKGs in ten important cancer signaling pathways with significant genetic variations. Most of the genes involved in these signaling pathways are CKGs. **G**) Two TCGA-analyzed signaling pathways of BLCA. FGFR3 and RPS6KA3, two control hubs, reside upstream of the most alternated genes in the pathways. The removal of edges from FGFR3 to RPS6KA3 will change their node types in the control scheme and therefore, make them sensitive CKGs.

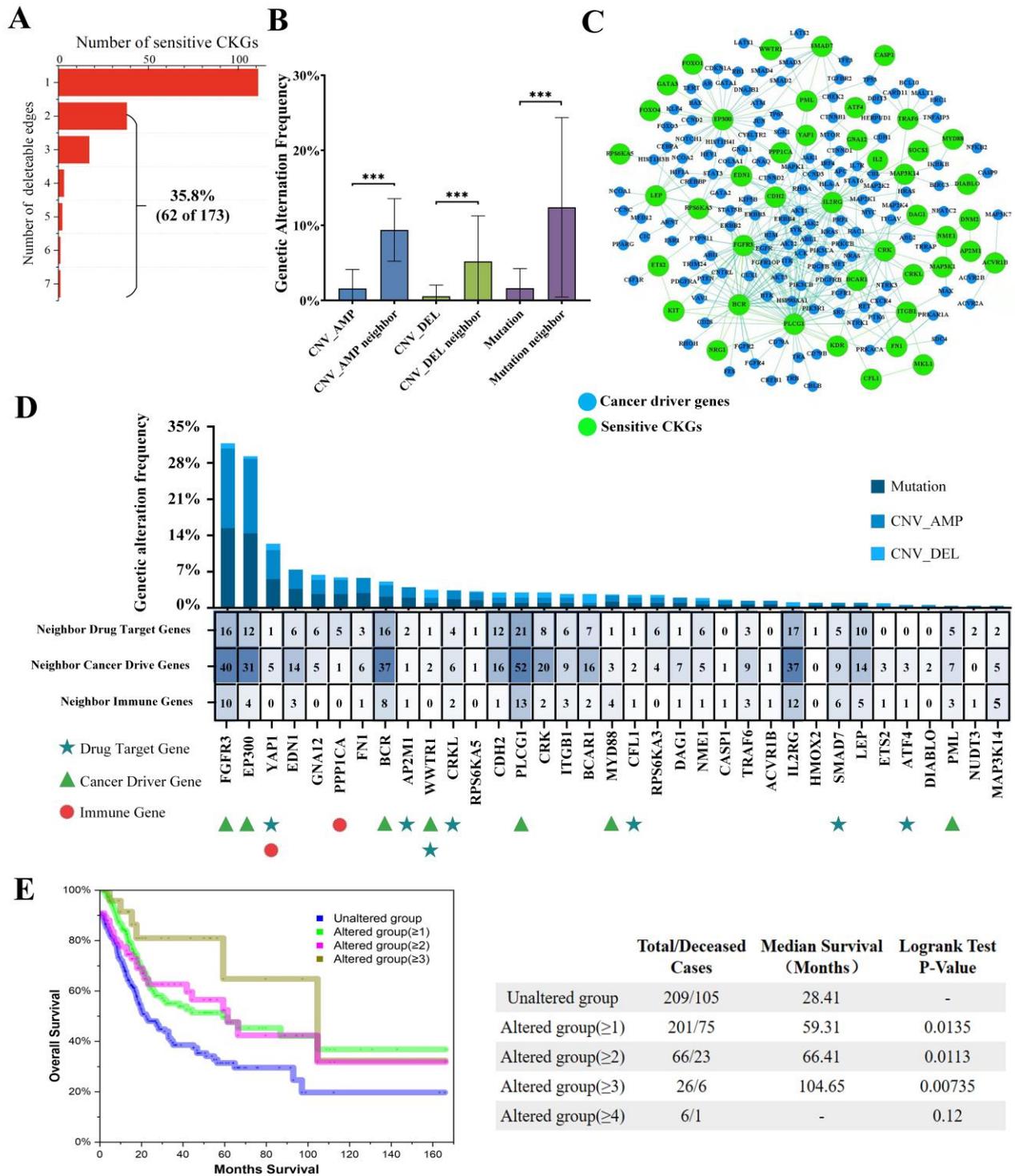

**Figure 3** | Characterizing sensitive CKGs in bladder cancer gene regulatory network. **A)** The distribution of the number of edges that could change a sensitive CKG. Most sensitive CKGs have less than three edges which could change their node types, indicating they are robust to the random structural perturbations. **B)** Differential analysis of the genetic alteration frequency between the sensitive CKGs and their neighbors. The genetic alteration frequency of sensitive CKGs is significantly lower than their neighbors by integrating the sample from the cBioPortal database. **C)** The subnetwork of sensitive CKGs and their surrounding cancer driver genes. **D)** Characterizing of 35 sensitive CKGs. Most of the 35 sensitive CKGs have low genetic alteration frequency in BLCA and most of them are directly connected to drug targets, cancer driver genes, and/or immune genes in BLCA_GRN. **E)** Survival analysis of the 35 sensitive CKGs in BLCA. The simultaneous

alteration of sensitive CKGs will significantly increase the survival rate of patients from 59.31 months (one sensitive CKGs altered) to 104.65 months (three or more sensitive CKGs altered).

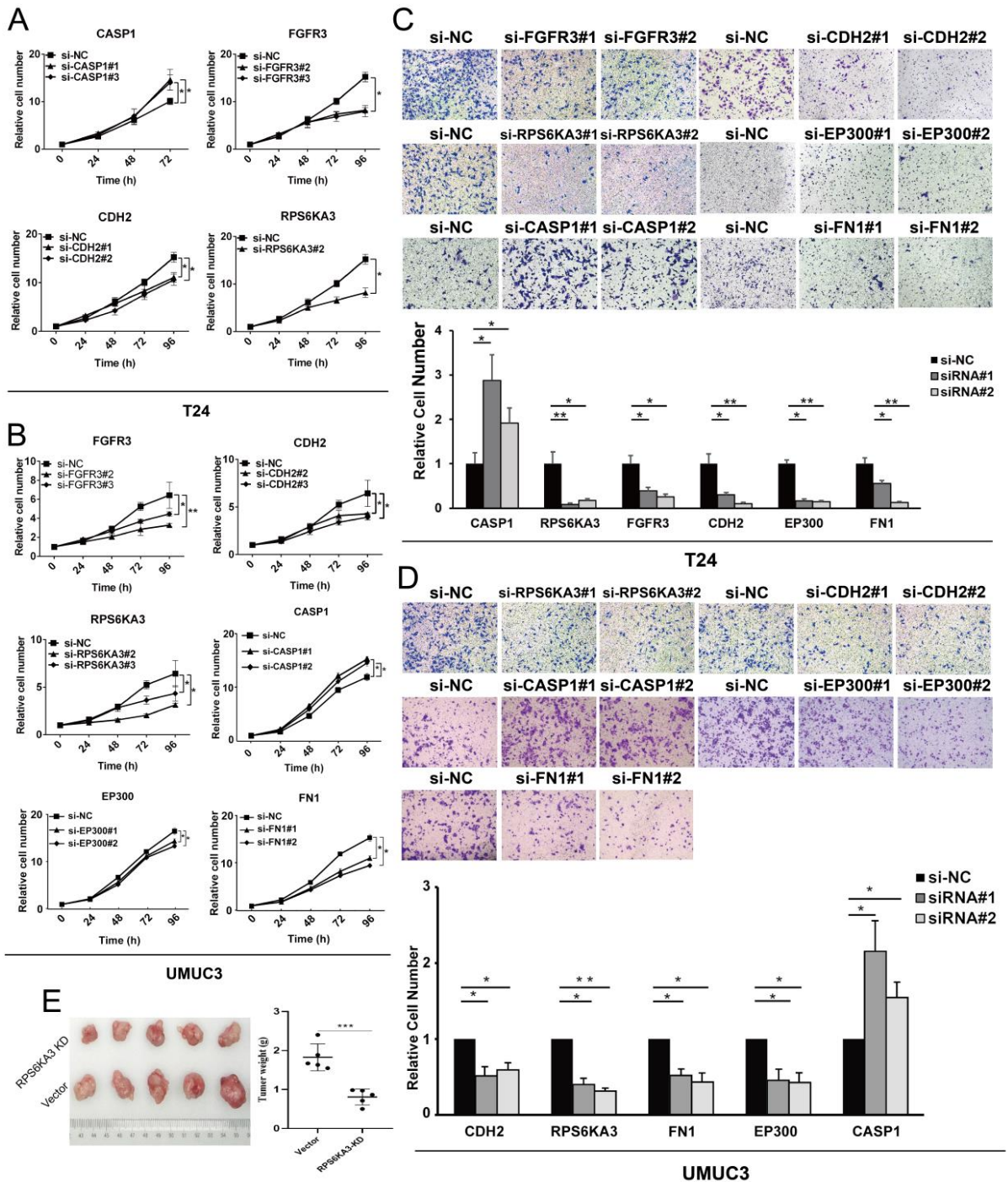

**Figure 4** | Experimental validation of representative sensitive CKGs in bladder cancer. **A-B)** CCK-8 assay and corresponding quantitative detection of the effect of siRNA-mediated knockdown sensitive CKGs on the proliferation of T24 (A) and UMUC3 (B) bladder cancer cells. **C-D)** Transwell migration assay(upper) and corresponding quantitative detection (down) of the effect of siRNA-mediated knockdown sensitive CKGs on the migration ability of T24 (C) and UMUC3 (D) bladder cancer cells. **E)** RPS6KA3 promoted tumor proliferation *in vivo*. BALB/c nude mice were injected with UMUC3 cells which were stably transfected with RPS6KA3 knockdown plasmids or control vector in the xenotransplantation model. After 4 weeks, it was found that the tumor size (left) and weight (right) in the RPS6KA3 knockdown group were significantly lower than those of the vector group. Data represent mean±s.d. from three replicate cultures. P-values were computed using the one-sided t-test.